\documentclass[a4paper,10pt]{article}
\usepackage[left=3.17cm,right=3.17cm,top=2.54cm,
headheight=0.5cm,headsep=0.54cm,bottom=2.54cm,footskip=0.79cm
]{geometry}

\usepackage{amsmath,amssymb,dsfont,cite}

\usepackage{graphicx,hyperref}

\usepackage{amsthm,float}

\newtheorem{prop}{Property}
\newtheorem{thm}{Theorem}

\begin{document}

\title{Curvature, area and Gauss-Bonnet formula of the Moyal sphere}

\author{Han-Liang Chen$^{\ast}$, Bing-Sheng Lin$^{\dag}$\\
	\small School of Mathematics, South China University of Technology,
	\small Guangzhou 510641, China\\
	\small $^{\ast}$Email: mj16661905@163.com\\
	\small $^{\dag}$Email: sclbs@scut.edu.cn\\
}

\date{\today}

\maketitle

\begin{abstract}
	We studied some geometric properties of the Moyal sphere. Using the conformal metric of the sphere in ordinary space and the matrix basis, we calculated the scalar curvature, total curvature integral and area of the Moyal sphere. We found that when the noncommutative parameter approaches to 0, the scalar curvature and area of the Moyal sphere return to those of the ordinary sphere. As the noncommutative parameter increases, the area of the Moyal sphere will decrease and eventually approach to 0. We found that the total curvature integral of the two-dimensional Moyal sphere still satisfies the usual Gauss-Bonnet formula and does not depend on the noncommutative parameter. We also calculated the approximate expression of the conformal metric with a constant curvature and obtained the corresponding correction function.
	In addition, we studied a type of generalized deformed Moyal sphere with two noncommutative parameters and obtained similar results. 
\end{abstract}

\maketitle

\section{Introduction}\label{sec1}
The ideas of noncommutative spacetime in physics started in 1947 \cite{Snyder}. In the 1980's, Connes formulated the mathematically rigorous framework of noncommutative geometry \cite{Connes}. Using the language of noncommutative algebras, noncommutative geometry provides new perspectives on other branches of mathematics, such as operator algebras, differential geometry, algebraic geometry, K-theory, cyclic cohomology, number theory, measure theory, etc. It also has many novel and useful applications in physics \cite{Seiberg, Douglas, Chaichian, Ettefaghi, lhc, Calmet, Couch, Muhuri, lh, Frob, Maris}. Many mathematicians have studied the generalizations of noncommutative analog of differential geometry from different perspectives \cite{Madore, Kaygun, Tretkoff, Fathizadeh, Dabrowski, Brzezinski, Arnlind, Floricel, Sciandra}. For example, Connes studied the geometric properties of noncommutative manifolds by studying the noncommutative form of the so-called spectral triples corresponding to Riemann manifolds. Some authors defined calculus on noncommutative algebras and then use the classical formula to define and calculate the scalar curvature on some noncommutative spaces \cite{Khalkhali,Wilson}.

Moyal space is one of the simplest noncommutative spaces, and has many important applications in mathematics and physics \cite{Gayral}. We will study some geometric properties of the Moyal sphere. As we all know, the calculations of geometric quantities on smooth manifolds depend on the operations of smooth functions on them, such as volume, curvature, vector field, differential form, tensor field, vector bundle, etc. The noncommutative geometric ideas based on this are described in detail in Ref.~\cite{Connes}. In Ref.~\cite{Eckstein}, the authors introduced the conformal metric $g=\frac{1}{h^{2}}(dx^{1}dx^{1}+dx^{2}dx^{2})$ into the two-dimensional Moyal space. Using the Moyal matrix basis and the orthonormal system method, they obtained the scalar curvature formula, and proved that there is a family of constant curvature metrics. The authors called the Moyal space with such constant curvature metrics the Moyal sphere. They also derived the same Gauss-Bonnet formula with the metric in a specific case as that of the ordinary two-dimensional sphere. Note that this is only one of the possible definitions of curvature in noncommutative spaces \cite{Eckstein}.

Since the classical sphere $\mathbf{S}^{n}$ is homeomorphic to $\mathbb{R}^{n}$ after removing a pole, and has a simple conformal metric $g=\frac{4}{(1+r^{2})^{2}}(dx^{1}dx^{1}+dx^{2}dx^{2}+\dots+dx^{n}dx^{n})$, one can use similar methods to generalize some results in Ref.~\cite{Eckstein} to high dimensional cases.
In this paper, we will study the generalizations of high-dimensional spheres in Moyal space from two different perspectives. On the one hand, we will study the $2M$-dimensional Moyal space with spherical metric $g=\frac{4A^{4}}{(r^{2}+A^{2})^{2}}(dx^{1}dx^{1}+dx^{2}dx^{2}+\dots+dx^{2M}dx^{2M})$.
This is slightly different from that in Ref.~\cite{Eckstein}. Based on this,  we will calculate the scalar curvature, area and total curvature integral of the Moyal spaces with certain dimensions.
On the other hand, similar to Ref.~\cite{Eckstein}, we will also study the Moyal space assigned a conformal metric with constant curvature analogous to that of the classical sphere. We will derive an approximate expression of the conformal factor function. These two type of noncommutative spaces can both return to the classical sphere when the noncommutative parameter approaches to 0, so they can all be regarded as the generalizations of ordinary sphere in Moyal space. We also call them \textit{Moyal spheres}.

It should be noted here that, in noncommutative spaces, there are no classical concepts of point and distance. So here the metric in Moyal space can only be considered as a generalized mathematical object analogous to that in classical space, and it does not have the usual intuitive geometric meaning.
A similar situation exists for concepts such as curvature, area and volume in noncommutative spaces. Of course, when the noncommutative parameter is set to $0$, these mathematical objects all return to the classical ones.

This paper is organized as follows: In Sec.~\ref{sec2}, we review the definitions and properties of Moyal star product and matrix basis. In Sec.~\ref{sec3}, we derive the expression for the curvature of Moyal space. In Sec.~\ref{sec4}, we calculate the scalar curvature of the two-dimensional Moyal sphere. In Sec.~\ref{sec5}, we calculate the Gauss-Bonnet formula of the two-dimensional Moyal sphere, and find that the result is the same as that of the ordinary two-dimensional sphere. This implies that the Moyal star product structure of the smooth function space does not change some topological properties of the two-dimensional sphere.
In Sec.~\ref{sec6} we also calculate the expression for scalar curvature of the four-dimensional Moyal sphere.
In Sec.~\ref{sec7}, we calculate the area of any even-dimensional Moyal sphere and a type of deformed Moyal sphere. We find that the area of Moyal sphere will decrease as the noncommutative parameter increases. When the noncommutative parameter approaches to infinity, the area of Moyal sphere will approach to 0. In Sec.~\ref{sec8}, we also study the conformal metric with constant curvature in the Moyal space, and give an approximate expression of the factor. The last section are the conclusions.

\section{Moyal star product and matrix basis}\label{sec2}
The Moyal $*$-product in the space of complex-valued functions on $\mathbb{R}^{2M}$ (where $M$ is a positive integer) is defined as \cite{Gayral}:
\begin{equation*}
	(f*g)(x):=(\pi \theta)^{-2M}\int_{\mathbb{R}^{2M}\times \mathbb{R}^{2M}}f(y)g(z)e^{\frac{2\mathbf{i}}{\theta}(x-y)\cdot\Lambda(x-z)}d^{2M}yd^{2M}z,
\end{equation*}
where the noncommutative parameter $\theta$\ is a real number, and the antisymmetric matrix
$\Lambda=I_{M}\otimes \varepsilon_2$, $I_{M}$\ is the $M\times M$ identity matrix, $\varepsilon_2$ is the antisymmetric matrix
$\varepsilon_2
=\begin{pmatrix}
	0&1 \\
	-1 & 0
\end{pmatrix}$.
The smooth function space $C^{\infty}(\mathbb{R}^{2M})$ forms an algebra with respect to the Moyal star product, called the Moyal algebra. The corresponding noncommutative space is the so-called Moyal space, and its coordinate variables $\{x_{1},x_{2},...,x_{2M}\}$ satisfy the commutation relations:
\begin{equation}
\left[x_{2j-1},x_{2j}\right]_*:=x_{2j-1}*x_{2j}-x_{2j}*x_{2j-1}
=-\left[x_{2j},x_{2j-1}\right]_*=\mathbf{i}\theta,\label{ms1}
\end{equation}
where $j=1,\dots,M$, and other commutators vanish.

A Schwarz space is a non-unitary, associative, involutive Fr\'echet algebra $(\mathcal{S} (\mathbb{R}^{2M})$,$*)$ with respect to the $*$-product, and it satisfies the following properties:
\begin{prop}
Let $f,g\in \mathcal{S} (\mathbb{R}^{2M},*) $, and $x=(x_{1},x_{2},...,x_{2M})$, there are\cite{Gayral}
\begin{itemize}
	\item[(i)]$(f*g)(x)\in \mathcal{S} (\mathbb{R}^{2M})$.
	\item[(ii)]When the variables of two functions are irrelevant with respect to the Moyal $*$-products, then their Moyal $*$-products return to the ordinary scalar products, for example: $f(x_{1},x_{2})*g(x_{3},x_{4})=f(x_{1},x_{2})g(x_{3},x_{4})$.
	\item[(iii)]$\partial^{j}(f*g)=\partial^{j}f*g+f*\partial^{j}g,~ j=1,2,...,2M$.
	\item[(iv)]$\int_{\mathbb{R}^{2M}}(f*g)(x)d^{2M}x=\int_{\mathbb{R}^{2M}}f(x)g(x)d^{2M}x$.
\end{itemize}
\end{prop}
Using the methods similar to those in Ref.~\cite{Eckstein}, one can introduce a set of functions on the Moyal space called the matrix basis $\left \{ f_{mn}\right \} _{m,n\in \mathbb{N}^{M}}$. First, let us consider a function in the Schwarz space $\mathcal{S}(\mathbb{R}^{2M}) $:
$f_{00}=2^{M}e^{-\frac{1}{\theta}\sum_{j=1}^{2M}x_{j}^{2}}$. Let $m=(m_{1},m_{2},...,m_{M})$,
and use multiple index notations: $ \left | m\right | =\sum_{j=1}^{M}m_{j}$, $m!=\prod_{j=1}^{M}m_{j} !$, $\delta_{ij}=\prod_{j=1}^{M}\delta_{m_{j}n_{j}} $. The matrix basis $\left \{ f_{mn}\right \} _{m,n\in \mathbb{N}^{M}}$ can be defined as:
\begin{equation*}
f_{mn}=\frac{1}{\sqrt{\theta^{\left | m\right |+\left | n\right |}m!n!}}\bar{a}^{m}*f_{00}*a^{n},
\end{equation*}
where $a^{n}=a_{1}^{n_{1}}...a_{M}^{n_{M}}$, the annihilation and creation functions are $a_{l}=\frac{1}{\sqrt{2}}(x_{l}+\mathbf{i}x_{l+M})$, $\bar{a}_{l}=\frac{1}{\sqrt{2}}(x_{l}-\mathbf{i}x_{l+M})$, respectively.
\begin{prop} The matrix basis $\left \{ f_{mn}\right \} _{m,n\in \mathbb{N}^{M}}$ satisfies the following properties:
\begin{itemize}
	\item[(i)]$f_{mn}*f_{kl}=\delta_{nk}f_{ml}$.
	\item[(ii)]$\bar{f}_{mn}=f_{nm}$.
	\item[(iii)]$\int_{\mathbb{R}^{2M}}f_{mn} d^{2M}x=(2\pi \theta)^{M}\delta_{mn}$.
\end{itemize}
\end{prop}
It is easy to see that $\left \{ f_{mn}\right \} _{m,n\in \mathbb{N}^{M}}$ is an orthogonal basis of the Schwarz space $\mathcal{S} (\mathbb{R}^{2M})$ under the $L^{2}$ norm. For the matrix basis on $\mathcal{S} (\mathbb{R}^{2})$, there are the following differential properties:
\begin{prop}
Consider the differential operators on $\mathcal{S} (\mathbb{R}^{2})$: $\partial=\frac{1}{\sqrt{2}}(\partial_{x_{1}}-\mathbf{i}\partial_{x_{2}}),\bar{\partial}=\frac{1}{\sqrt{2}}(\partial_{x_{1}}+\mathbf{i}\partial_{x_{2}})$, we have\cite{Eckstein}
\begin{align*}
	\partial f_{mn} &= \sqrt{\frac{n}{\theta}}f_{m,n-1}-\sqrt{\frac{m+1}{\theta}}f_{m+1,n}, \\
	\bar{\partial}f_{mn} &= \sqrt{\frac{m}{\theta}}f_{m-1,n}-\sqrt{\frac{n+1}{\theta}}f_{m,n+1}.
\end{align*}
\end{prop}
The above differential formulas are valid for $m,n \ge 0$ (when a subscript is less than $0$, the term is set to be $0$). It is easy to obtain the following partial differential equations and the formula acted by the Laplace operator of the matrix basis:
\begin{align*}
\partial_{x_{1}}f_{mn} &=\frac{1}{\sqrt{2}}(\partial+\bar{\partial})f_{mn}\\
&=\frac{1}{\sqrt{2}}\left( \sqrt{\frac{n}{\theta}}f_{m,n-1}- \sqrt{\frac{m+1}{\theta}}f_{m+1,n}+\sqrt{\frac{m}{\theta}}f_{m-1,n}-\sqrt{\frac{n+1}{\theta}}f_{m,n+1}\right), \\
\partial_{x_{2}}f_{mn} &=\frac{1}{\sqrt{2}\mathbf{i}}(\bar{\partial}-\partial)f_{mn}\\
&= \frac{1}{\sqrt{2}\mathbf{i}}\left( \sqrt{\frac{n}{\theta}}f_{m,n-1}-\sqrt{\frac{m+1}{\theta}}f_{m+1,n}-\sqrt{\frac{m}{\theta}}f_{m-1,n}+\sqrt{\frac{n+1}{\theta}}f_{m,n+1}\right) , \\
\triangle f_{mn}
&=2\partial\bar{\partial}f_{mn}\\
&=\frac{2}{\theta}\left[-(m+n+1)f_{mn}+\sqrt{(m+1)(n+1)}f_{m+1,n+1 }+\sqrt{mn}f_{m-1,n-1}\right].
\end{align*}
In addition, the matrix basis $\{f_{mm}\}_{m=0}^{\infty}$ also satisfies the following very useful formulas \cite{Eckstein}:
\begin{equation}\label{fr}
\sum_{m}f_{mm}=1 ,\qquad \sum_{m}mf_{mm}=\frac{1}{2}\left(\frac{r^{2}}{\theta}-1 \right),
\end{equation}
where $r=\sqrt{x_1^2+x_2^2}$.

\section{Curvature of the Moyal space}\label{sec3}
Similar to Ref.~\cite{Eckstein}, we will
consider the conformal metric $g=h^{-2}(dx^{1}dx^{1}+dx^{2}dx^{2}+ \dots dx^{n}dx^{n})$ for an $n$-dimensional Riemannian manifold $M$, where $h$ is some smooth positive function, and $n\geq 2$. By virtue of the method used in Ref.~\cite{Dabrowski}, here we will choose a set of local orthonormal bases $\left\{e_{i}=h\frac{\partial}{\partial x^{i}}\right\}$ in its local coordinate system. Using the properties of the Levi-Civita connection $\nabla$, one can obtain the following expression of the curvature tensor:
\begin{align*}
	R(e_{i},e_{j},e_{i},e_{j})&=\langle R(e_{i},e_{j})e_{i},e_{j}\rangle\\
	&=\langle\nabla_{e_{j}}\nabla_{e_{i}}e_{i}-\nabla_{e_{i}}\nabla_{e_{j}}e_{i}+\nabla_{[ e_{i},e_{j}]}e_{i},e_{j}\rangle \\
	&=e_{j}c_{jii}+e_{i}c_{ijj}-\frac{1}{4}\sum_{k}(c_{ kji}+c_{jik}+c_{kij})(c_{jik}+c_{ikj}+c_{jki})\\
	&\qquad-\sum_{k}c_{kii}c_{kjj}+ \frac{1}{2}\sum_{k}c_{ijk}(c_{jki}+c_{kij}+c_{jik}),
\end{align*}
where $R(X,Y,Z,W)$ is the Riemann curvature tensor, $\langle \cdot ,\cdot \rangle$ is the Riemann metric of the manifold, and $c_{ijk}$ are the structure constants of the commutators $[e_{i},e_{j}]=\displaystyle \sum_{k=1}^{n} c_{ijk}e_{k}$.
The scalar curvature $S$ is:
\begin{align*}
	S&=\sum_{i,j=1}^{n}R(e_{i},e_{j},e_{i},e_{j})\\
	&=\sum_{i,j=1}^{n}\bigg[e_{j}c_{jii}+e_{i}c_{ijj}-\sum_{k}c_{kii}c_{kjj }-\frac{1}{4}\sum_{k}(c_{kji}+c_{jik}+c_{kij})(c_{jik}+c_{ikj}+c_{jki})\\
	&\qquad +\frac{1}{2}\sum_{k}c_{ijk}(c_{jki}+c_{kij}+c_{jik})\bigg]\\
	&=2\sum_{ij}^{n}e_{i}c_{ijj}-\sum_{ijk}^{n}c_{kii}c_{kjj}-\frac{1}{2}\sum_{ijk}^{n}c_{ijk}c_{ikj}-\frac{1}{4}\sum_{ijk}^{n}c_{ijk}c_{ijk}.
\end{align*}
Considering the symmetry of the summation index, and using the Einstein summation convention, one can obtain the following scalar curvature formula:
\begin{equation*}
	S=2e_{i}c_{ijj}-c_{kii}c_{kjj}-\frac{1}{4}c_{ijk}c_{ijk}-\frac{1}{2}c_{ijk}c_{ikj}.
\end{equation*}
This is just the same as the result obtained in Ref.~\cite{Dabrowski}.
In the Moyal space, ordinary multiplications between functions will be replaced by Moyal $*$-products. Similar to the result in Ref.~\cite{Eckstein}, there is the scalar curvature in the Moyal space:
\begin{equation}\label{scf}
	S_\theta=2e_{i}c_{ijj}-c_{kii}*c_{kjj}-\frac{1}{4}c_{ijk}*c_{ijk}-\frac{1}{2}c_{ijk}*c_{ikj}.
\end{equation}
Due to the symmetry of the sum, adding the $*$-product does not cause any confusion in the definition.

Note that the scalar curvature (\ref{scf}) only contains the structure constants and their derivatives. So in order to obtain the scalar curvature, one only need to calculate the structure constants under the set of orthonormal basis :
\begin{align*}
	[e_{i},e_{j}]_*&=h*\partial_{i}(h*\partial_{j})-h*\partial_{j}(h*\partial_{i})\\
	&=h*\partial_{i}(h)*h_*^{-1}*(h*\partial_{j})-h*\partial_{j}(h)*h_*^{-1}*(h*\partial_{i})\\
	&=h*\partial_{i}(h)*h_*^{-1}*e_{j}-h*\partial_{j}(h)*h_*^{-1}*e_{i}\\
	&=\sum_{k=1}^{n} c_{ijk}e_{k},
\end{align*}
where $h_*^{-1}$ is the inverse of $h$ with respect to the Moyal $*$-product, $h*h_*^{-1}=h_*^{-1}*h=1$.
Note that $c_{ijk}\ne 0$ only when $i\ne j$ and $k=i$ or $j$, and we have
$$c_{iji}=-h*\partial_{j}(h)*h_*^{-1}, 
\qquad c_{ijj}=h*\partial_{i}(h)*h_*^{-1} \qquad(i\ne j),$$
and
$$
e_{i}c_{ijj}=h*\partial_{i}(h*\partial_{i}(h)*h_*^{-1})
=h*\partial_{i}(h)_*^{2}*h_*^{-1}+h_*^{2}*\partial_{ii}(h)*h_*^{-1}-h_*^{2}*\partial_{i}(h)*h_*^{-1}*\partial_{i}(h)*h_*^{-1}.
$$
Substituting the above expressions into Eq.(\ref{scf}), one can get the expression for scalar curvature of the $n$-dimensional Moyal space:
\begin{align}\label{sss}
	S_\theta&=2(n-1)h_*^{2}*\partial_{ii}(h)*h_*^{-1}-(n^{2}-3n+2)h*\partial_{i}(h)_*^{2}*h_*^{-1}\nonumber\\
	&\quad -2(n-1)h_*^{2}*\partial_{i}(h)*h_*^{-1}*\partial_{i}(h)*h_*^{-1}.
\end{align}
Here we have used the Einstein summation convention, and the powers and inverses of functions in the above equation are in the sense of Moyal $*$-product. It is easy to see that using the metric $g=h_*^{-2}(dx^{1}dx^{1}+dx^{2}dx^{2}+...+dx^{n}dx^{n}) $, the generalization of the classical scalar curvature formula (\ref{scf}) is well defined (because the indices of the summation terms are symmetric, after commuting the star products of the two summation terms in the sum, the results are still the same).

\section{Scalar curvature of the two-dimensional Moyal sphere}\label{sec4}
Let us recall that in ordinary space, the round metric of a sphere with a radius of $A$ in stereographic coordinates is:
\begin{equation*}
	g=\frac{4A^4}{(A^2+r^{2})^{2}}(dx^{1}dx^{1}+dx^{2}dx^{2}).
\end{equation*}
Therefore, the metric $g$ can be written as:
\begin{equation*}
	g=\frac{1}{h^{2}}(dx^{1}dx^{1}+dx^{2}dx^{2}),
\end{equation*}
where
$$h=\frac{A^2+r^{2}}{2A^2}=\sum_{m}\phi_{m}f_{mm},\qquad \phi_{m}=\frac{2\theta m+A^2+\theta}{2A^2},\qquad m=0,1,\dots .$$
As the other conformally rescaled metrics, one can use this function $h$ to rescale the orthonormal frames in Moyal space. 
Now we will use this special expression of $h$ into the formula (\ref{sss}) to derive the scalar curvature of two dimensional Moyal sphere.

After some straightforward calculations, one can obtain
\begin{equation*}
	\triangle h=\frac{2}{\theta}\sum_{m}[-(2m+1)\phi_{m}+m\phi_{m-1}+(m+1)\phi_{m+1}]f_{mm},
\end{equation*}
\begin{equation*}
	\partial_{i}(h)*h_*^{-1}*\partial_{i}(h)
	=\frac{1}{\theta}\sum_{m=0}\left[ \frac{m}{\phi_{m-1}}(\phi_{m}-\phi_{m-1})^{2}+\frac{m+1}{\phi_{m+1}}(\phi_{m+1}-\phi_{m})^{2}\right] f_{mm}.
\end{equation*}
Substituting the above results into Eq.(\ref{sss}), one can get the scalar curvature of the two-dimensional Moyal sphere
\begin{eqnarray}
	S_\theta&=&2[h_*^{2}*\partial_{ii}(h)*h_*^{-1}-h_*^{2}*\partial_{i}(h)*h_*^{-1}*\partial_{i}(h)*h_*^{-1}]\nonumber\\
	&=&\frac{2}{\theta}\sum_{m}\phi_{m}\left[ \frac{m+1}{\phi_{m+1}}(\phi_{m+1}^{2}-\phi_{m}^{2})-\frac{m}{\phi_{m-1}}(\phi_{m}^{2}-\phi_{m-1}^{2})\right] f_{mm}\nonumber\\
	&=&\sum_{m}\left[ \frac{2}{A^{2}}-\frac{2\theta(\theta+A^{2})}{A^{4}(2\theta m+A^{2}+3\theta)}+\frac{2\theta(A^{2}-\theta)}{A^{4}(2\theta m+A^{2}-\theta)}\right] f_{mm}\nonumber\\
	&=&\frac{2}{A^{2}}-\frac{2\theta(A^{2}+\theta)}{A^{4}}(r^{2}+A^{2}+2\theta)_{*}^{-1}+\frac{2\theta(A^{2}-\theta)}{A^{4}}(r^{2}+A^{2}-2\theta)_{*}^{-1}.\label{sth}
\end{eqnarray}
In the last equation, we have used the relations (\ref{fr}).
Note that $S_\theta=S_{-\theta}$, without losing generality, one can be assumed that $\theta>0$.
Usually, we consider $\theta$ to be a very small number relative to $A$, especially from a physical point of view. For example, in the quantum phase space, the corresponding noncommutative parameter is just the Planck constant, which is extremely small.
On the other hand, in the above calculations, because of the expressions for the inverses with respect to the Moyal product, when $\theta$ takes some values, there will be some singularities.
In order to avoid singularities, one can choose $\theta\in (0,\frac{A^2}{2})$. Obviously, when $\theta \to 0$, there is $S_\theta\to \frac{2}{A^2}$, which is the same as the scalar curvature of the sphere $\mathbf{S}^{2}$ with a radius of $A$ in the classical case.

When the noncommutative parameter $\theta$ is very small, using the methods similar to those in Ref.~\cite{Eckstein}, one can only consider terms that approximate to the $\theta^2$ term. The inverse of a radial function $f$ with respect to the Moyal $*$-product has an approximate expansion:
$$f_{*}^{-1}=f^{-1}+\frac{\theta^{2}}{4r}[(f')^{3}f^{-4}-f''f'f^{-3}]+o(\theta^{2}).$$
Therefore, the scalar curvature of the Moyal sphere (\ref{sth}) can be approximately expressed as:
\begin{align*}
	S_\theta&\approx  \frac{2}{A^{2}}-\frac{2\theta(A^{2}+\theta)}{A^{4}(r^{2}+A^{2}+2\theta)}
	+\frac{2\theta(A^{2}-\theta)}{A^{4}(r^{2}+A^{2}-2\theta)}\\
	&= \frac{2}{A^{2}}\cdot \eta_\theta(r;\lambda),
\end{align*}
where $\lambda=\frac{A^{2}}{\theta}$, and
$$
\eta_\theta(r;\lambda)=1-\frac{\lambda+1}{\lambda\left( \frac{r^{2}}{\theta}+\lambda+2\right) }
+\frac{\lambda-1}{\lambda\left( \frac{r^{2}}{\theta}+\lambda-2\right) }.
$$
\begin{figure}[H]
	\centering
	\includegraphics[scale=0.6]{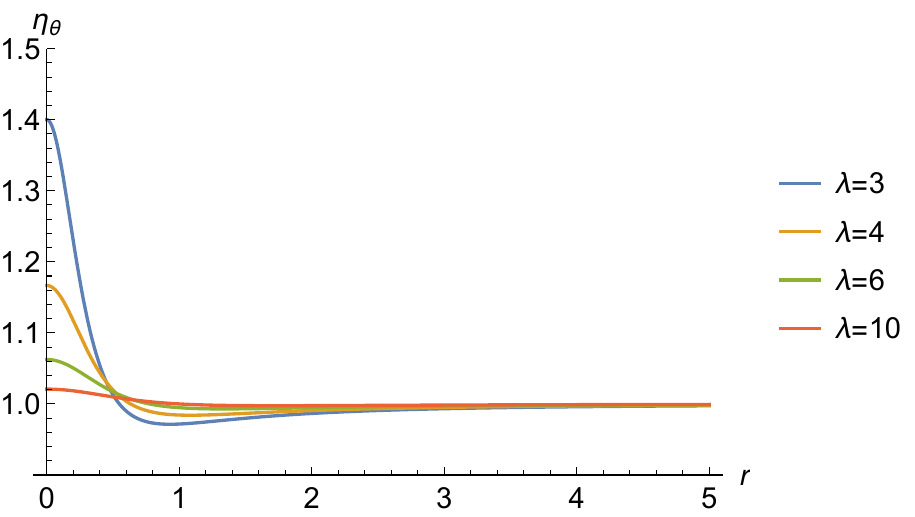}
	\caption{The curves of coefficient $\eta_\theta(r;\lambda)$ with different values of $\lambda$ ($\theta=0.1$). }\label{fig1}
\end{figure}
In Fig.~\ref{fig1}, we show the curves of coefficient $\eta_\theta(r;\lambda)$ with different values of $\lambda$, where the noncommutative parameter $\theta=0.1$.
One can see that for a given small $\theta$, when $\lambda \to \infty$, or equivalently, when the radius $A\to \infty$, the scalar curvature of the Moyal sphere will approach to that of the ordinary sphere. When $\lambda \to 0$, or equivalently, when the radius $A\to 0$, the scalar curvature of the Moyal sphere is significantly different from that of the ordinary sphere for small $r$. But when $r$ is very large, the result also approaches to that of the ordinary sphere.

\section{Gauss-Bonnet formula for the Moyal sphere}\label{sec5}
\begin{thm}[Gauss-Bonnet formula for compact surfaces]
	Let $M$ be a compact, oriented two-dimensional surface, there is
	\begin{equation*}
		\int_{M}Kd\sigma=2\pi \chi(M),
	\end{equation*}
	where $K$ is the Gaussian curvature of $M$, $d\sigma$ is the area element, and $\chi(M)$ is the Euler characteristic number of $M$.
\end{thm}
Since the scalar curvature $S=2K$ in two-dimensional surfaces, the Gauss-Bonnet formula can be written as:
\begin{equation*}
	\int_{M}Sd\sigma=4\pi \chi(M).
\end{equation*}
In particular, the Euler characteristic number of the two-dimensional sphere $\mathbf{S}^{2}$ is $\chi(\mathbf{S}^{2})=2$, and $\int_{\mathbf{S}^{2}}Sd\sigma=8\pi$.

Now let us calculate the Gauss-Bonnet formula for the two-dimensional Moyal sphere.
\begin{align*}
	\int_{\mathbf{S}^{2}}S_\theta*d\sigma
	&=\int_{\mathbb{R}^{2}}\sum_{m}\left[ \frac{2}{A^{2}}-\frac{2\theta(\theta+A^{2})}{A^{4}(2\theta m+A^{2}+3\theta)}+\frac{2\theta(A^{2}-\theta)}{A^{4}(2\theta m+A^{2}-\theta)}\right] f_{mm}*h_{*}^{-2} dxdy\\
	&=\int_{\mathbb{R}^{2}}\left\{ \sum_{m}\left[ \frac{2}{A^{2}}-\frac{2\theta(\theta+ A^{2})}{A^{4}(2\theta m+A^{2}+3\theta)}+\frac{2\theta(A^{2}-\theta)}{A^{4}(2\theta m+A^{2}-\theta)}\right] f_{mm}\right. \\
	&\qquad\left. *\sum_{m}\frac{4A^{4}}{(2\theta m +A^{2}+\theta)^2}f_{mm}\right\} dxdy\\
	&=\int_{\mathbb{R}^{2}}\sum_{m}\frac{8(2\theta A^{2}m+A^{4}-2\theta^{2}+\theta A^{2})}{(2\theta m+A^{2}+\theta)(2\theta m+A^{2}-\theta)(2\theta m+A^{2}+3\theta)}f_{mm}dxdy\\
	&=2\pi\theta\sum_{m}\frac{8(2\theta A^{2}m+A^{4}-2\theta^{2}+\theta A^{2})} {(2\theta m+A^{2}+\theta)(2\theta m+A^{2}-\theta)(2\theta m+A^{2}+3\theta)}\\
	&=4\pi\sum_{m}\left[(A^2-\theta)B_m+2\theta B_{m+1}-(A^2+\theta) B_{m+2}\right],
	\end{align*}
where
$$
B_m=\frac{1}{2m\theta+A^{2}-\theta}.
$$
So we have
\begin{align*}	
	\int_{\mathbf{S}^{2}}S_\theta*d\sigma
	&=4\pi\left[\sum_{m=0}^{\infty}(A^2-\theta)(B_m-B_{m+1})+\sum_{m=0}^{\infty}(A^2+\theta)(B_{m+1}-B_{m+2}) \right]\\
	&=4\pi\lim_{m\to \infty}\left[(A^2-\theta)(B_0-B_{m+1})+(A^2+\theta)(B_{1}-B_{m+2}) \right]\\
	&=4\pi\left[(A^2-\theta)B_0+(A^2+\theta)B_{1}\right]\\
	&=8\pi.
	\end{align*}
The above results show that the Gauss-Bonnet formula for the two-dimensional Moyal sphere is independent of the value of the noncommutative parameter $\theta$. It is the same as that of the classical two-dimensional sphere $\mathbf{S}^{2}$ with radius $A$.

Therefore, the Gauss-Bonnet integrals for two-dimensional Moyal spheres of different radii are all equal to $8\pi$.
This means that the integral of the scalar curvature on the surface in the noncommutative sense is only related to the topological invariant (Euler characteristic number) of the surface, and has nothing to do with the noncommutative parameter $\theta$. This result is consistent with that in Ref.~\cite{Eckstein}. But it should be noted that an appropriate $\theta \in (0,\frac{A^2}{2})$ must be chosen here so that the inverse with respect to the Moyal star product of the matrix basis series exists.

\section{Scalar curvature of the four-dimensional Moyal sphere}\label{sec6}
Considering the conformal metric $g=h^{-2}(dx^{1}dx^{1}+dx^{2}dx^{2}+dx^{3}dx^{3}+dx^{4}dx^{4})$ of the 4-dimensional sphere, one can similarly assume
$$h=\sum_{m}\phi_{m}f_{mm}(x_1,x_2)+\sum_{n}\psi_{n}g_{nn}(x_3,x_4)=\sum_{m,n=0}(\phi_{m}+\psi_{n})f_{mm}g_{nn}.
$$
After some straightforward calculations, one can obtain
$$
	h_*^{-1}*\partial_{i}(h)_*^{2}=\frac{1}{\theta}\sum_{m,n=0}\frac{1}{\phi_{m}+\psi_{n}}[m(\phi_{m}-\phi_{m-1})^{2}+(m+1)(\phi_{m+1}-\phi_{m})^{2}]f_{mm}g_{nn}.
$$
The scalar curvature of the four-dimensional Moyal space (\ref{sss}) is
\begin{align*}
	S_\theta&=6[h_*^{2}*\partial_{ii}(h)*h_*^{-1}-h*\partial_{i}(h)_*^{2}*h_*^{-1}-h_*^{2}*\partial_{i}(h)*h_*^{-1}*\partial_{i}(h)*h_*^{-1}]\\
	&=\frac{6}{\theta}\sum_{m,n=0}(\phi_{m}+\psi_{n})\Bigg\{[-(4m+2)\phi_{m}+2m\phi_{m-1}+2(m+1)\phi_{m+1}]\\
	&\qquad-\bigg[m(\phi_{m}-\phi_{m-1})^{2}\bigg(\frac{1}{\phi_{m}+\psi_{n}}+\frac{1}{\phi_{m-1}+\psi_{n}}\bigg)\\
	&\quad\qquad+(m+1)(\phi_{m+1}-\phi_{m})^{2}\bigg(\frac{1}{\phi_{m}+\psi_{n}}+\frac{1}{\phi_{m+1}+\psi_{n}}\bigg)\bigg]\Bigg\}f_{mm}g_{nn}\\
	&\quad+\frac{6}{\theta}\sum_{m,n=0}(\phi_{m}+\psi_{n})\Bigg\{[-(4n+2)\psi_{n} +2n\psi_{n-1}+2(n+1)\psi_{n+1}]\\
	&\qquad-\bigg[n(\psi_{n}-\psi_{n-1})^{2}\bigg(\frac{1}{\phi_{m}+\psi_{n}}+\frac{1}{\psi_{n-1}+\phi_{m}}\bigg)\\
	&\quad\qquad+(n+1)(\psi_{n+1}-\psi_{n})^{2}\bigg(\frac{1}{\phi_{m}+\psi_{n}} +\frac{1}{\psi_{n+1}+\phi_{m}}\bigg)\bigg]\Bigg\}f_{mm}g_{nn}.
\end{align*}
For the conformal metric of the four-dimensional sphere with radius $A$, we have
$$
h=\frac{r^{2}+A^2}{2A^2}=\sum_{m,n}\frac{2\theta (m+n)+2\theta+A^2}{2A^2}f_{mm}g_ {nn}.
$$
Therefore, one can take $\phi_{m}=\frac{4\theta m+2\theta+A^2}{4A^2} ,\psi_{n}=\frac{4\theta n+2\theta+A^2}{4A^2}$. So we have
	\begin{align*}
	S_{\theta}=&\sum_{m, n=0}\left[\frac{12}{A^{2}}+\frac{6\theta }{A^{2}\left(A^{2}+2\theta(m+n)\right)}-\frac{6\theta }{A^{2}\left(4\theta+A^{2}+2\theta(m+n)\right)}\right]f_{mm}g_{nn}\\
	=&\frac{12}{A^{2}}+\frac{6\theta}{A^{2}}\left[(\rho^{2}+\sigma^{2}-2\theta+A^{2})_{*}^{-1}-(\rho^{2}+\sigma^{2}+2\theta+A^{2})_{*}^{-1}\right]\\
	\approx &\frac{12}{A^{2}}+\frac{6\theta}{A^{2}\left(r^{2}+A^{2}-2\theta\right)}-\frac{6\theta}{A^{2}\left(r^{2}+A^{2}+2\theta\right)}\\
	=&\frac{12}{A^{2}}\Lambda_{\theta}(r;\lambda),
\end{align*}
where $\rho^{2}=x_{1}^{2}+x_{2}^{2}$, $\sigma^{2}=x_{3}^{2}+x_{4}^{ 2}$, $r=\sqrt{x_{1}^{2}+x_{2}^{2}+x_{3}^{2}+x_{4}^{2}}$, and
$$
\Lambda_{\theta}(r;\lambda)=1 +\frac{\theta}{2\left(r^{2}+\lambda\theta-2\theta\right)}-\frac{\theta}{2\left(r^{2}+\lambda\theta+2\theta\right)}.
$$
The calculations of the above equation approximates to the second-order term of $\theta$.

\begin{figure}[H]
	\centering
	\includegraphics[scale=0.6]{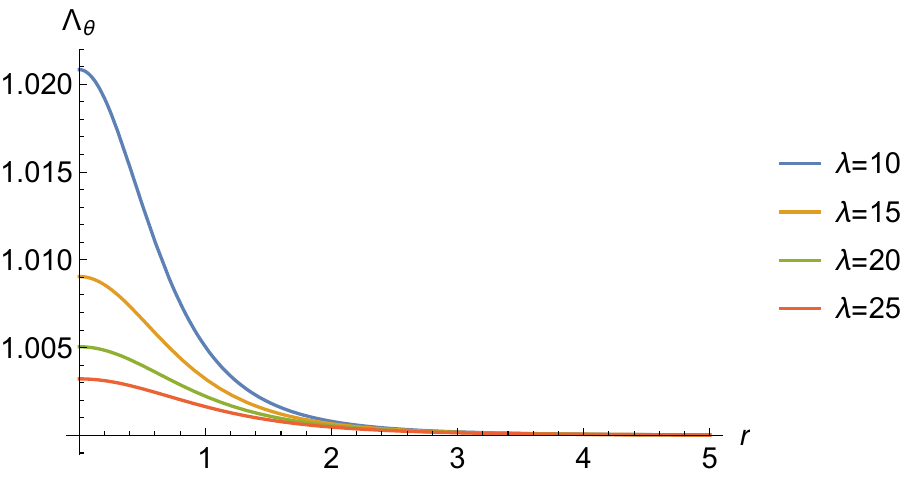}
	\caption{The curves of coefficient $\Lambda_\theta(r;\lambda)$ with different values of $\lambda$ ($\theta=0.1$). }\label{fig2}
\end{figure}

Similarly, in order to avoid singularities, one can choose $0<\theta <\frac{A^{2}}{2}$. Obviously, when $\theta \to 0$, there is $S_\theta\to \frac{12}{A^2}$. This is also the scalar curvature of the classical four-dimensional sphere.
Taking $\theta=0.1$, the values of coefficient $\Lambda_{\theta}(r;\lambda)$ are shown in Fig.~\ref{fig2}.
From Fig.~\ref{fig2}, one can see that for a given small $\theta$, when $\lambda \to \infty$, or equivalently, when the radius $A\to \infty$, the scalar curvature of the four-dimensional Moyal sphere will approach to that of the ordinary sphere. When $\lambda \to 0$, or equivalently, when the radius $A\to 0$, the scalar curvature of the Moyal sphere is different from that of the ordinary sphere for small $r$. However, they are all very close to $\frac{12}{A^2}$, which is the curvature of the classic four-dimensional sphere. When $r$ is very large, the result also approaches to that of the ordinary sphere. This is similar to the result of the two-dimensional case.

\section{Area of the Moyal sphere}\label{sec7}
In general, the conformal metric of the $2M$-dimensional Moyal sphere can be expressed as $g=h_{*}^{-2}(dx^{1}dx^{1}+\dots+dx^{2M}dx^{2M})$,
where
$$
h=\frac{r^{2}+A^{2}}{2A^{2}}
=\sum_{m_{1},m_{2},\dots,m_{M}}\frac{2\theta (m_{1}+m_{2}+\dots m_{M})+A^{2}+M\theta}{2A^{2}}f_{m_{1}m_{1}}f_{m_{2}m_{2}}\dots f_{m_{M}m_{M}}.
$$
By virtue of the formula for the classical sphere, one can calculate the area of the $2M$-dimensional Moyal sphere:
\begin{align}\label{s2m}
	\int_{\mathbf{S}^{2M}}dvol_{*}&=\int_{\mathbb{R}^{2M}}h_{*}^{-2M}d^{2M}x\nonumber\\
	&=\int_{\mathbb{R}^{2M}}\sum_{m_{1},m_{2},\dots,m_{M}}\frac{(2A^{2})^{2M}}{\left[2\theta (m_{1}+m_{2}+\dots m_{M})+A^{2}+M\theta\right]^{2M}}f_{m_{1}m_{1}}f_{m_{2}m_{2}}\dots f_{m_{M}m_{M}}d^{2M}x\nonumber\\
	&=\left(2\pi \theta\right)^{M}\sum_{m_{1},m_{2},\dots,m_{M}}\frac{(2A^{2})^{2M}}{\left[2\theta (m_{1}+m_{2}+\dots m_{M})+A^{2}+M\theta\right]^{2M}}\nonumber\\
	&=\frac{(2\pi A^{4})^{M}}{\theta^{M}}\sum_{k=1}^{\infty}\frac{C_{k-2+M}^{M-1}}{\left(k+\frac{A^{2}}{2\theta}+\frac{M}{2}-1\right) ^{2M}}\nonumber\\
	&=\frac{2\pi^{M+\frac{1}{2}}A^{2M}}{\Gamma\left(M+\frac{1}{2}\right) }\cdot\frac{2^{M-1}\lambda^{M}}{\sqrt{\pi}}\Gamma\left(M+\frac{1}{2}\right)\sum_{k=0}^{\infty}\frac{C_{k-1+M}^{M-1}}{\left(k+\frac{\lambda}{2}+\frac{M}{2}\right)^{2M}}\nonumber\\
	&=\frac{2\pi^{M+\frac{1}{2}}A^{2M}}{\Gamma\left(M+\frac{1}{2}\right) }\cdot\gamma_{M}(\lambda),		
\end{align}
where $\Gamma(x)$ is the Gamma function, $C^{M}_{k}=\frac{k!}{M!(k-M)!}$, and
\[
\gamma_{M}(\lambda)=\frac{2^{M-1}\lambda^{M}}{\sqrt{\pi}}\Gamma\left(M+\frac{1}{2}\right)\sum_{k=0}^{\infty}\frac{C_{k-1+M}^{M-1}}{\left( k+\frac{\lambda}{2}+\frac{M}{2}\right)^{2M}}.
\]
For a given $M$, $\gamma_{M}(\lambda)$ can also be expressed as a sum of polygamma functions. For example, for $M=1, 2$, we have
\[
\gamma_{1}(\lambda)=\frac{\lambda}{2} \Psi^{(1)}\left(\frac{\lambda+1}{2}\right),\qquad
\gamma_{2}(\lambda)=\frac{\lambda^{2}}{8}\left[-6\Psi^{(2)}\left(1+\frac{\lambda}{2}\right)-\lambda\Psi^{(3)}\left(1+\frac{\lambda}{2}\right)\right],
\]
where $\Psi^{(n)}(x)$ is a polygamma function.
Therefore, the areas of the 2-dimensional and 4-dimensional Moyal spheres are
\[
\int_{\mathbf{S}^{2}}dvol_{*}=4\pi A^{2}\cdot \gamma_{1}(\lambda)
=2\pi A^{2}\lambda \Psi^{(1)}\left(\frac{\lambda+1}{2}\right),
\]
\[
\int_{\mathbf{S}^{4}}dvol_{*}=\frac{8\pi^{2}A^{4}}{3}\cdot\gamma_{2}(\lambda)
=\frac{\pi^{2}A^{4}\lambda^{2}}{3}\left[-6\Psi^{(2)}\left(1+\frac{\lambda}{2}\right)-\lambda\Psi^{(3)}\left(1+\frac{\lambda}{2}\right)\right].
\]
\begin{figure}[H]
\centering
\includegraphics[scale=0.27]{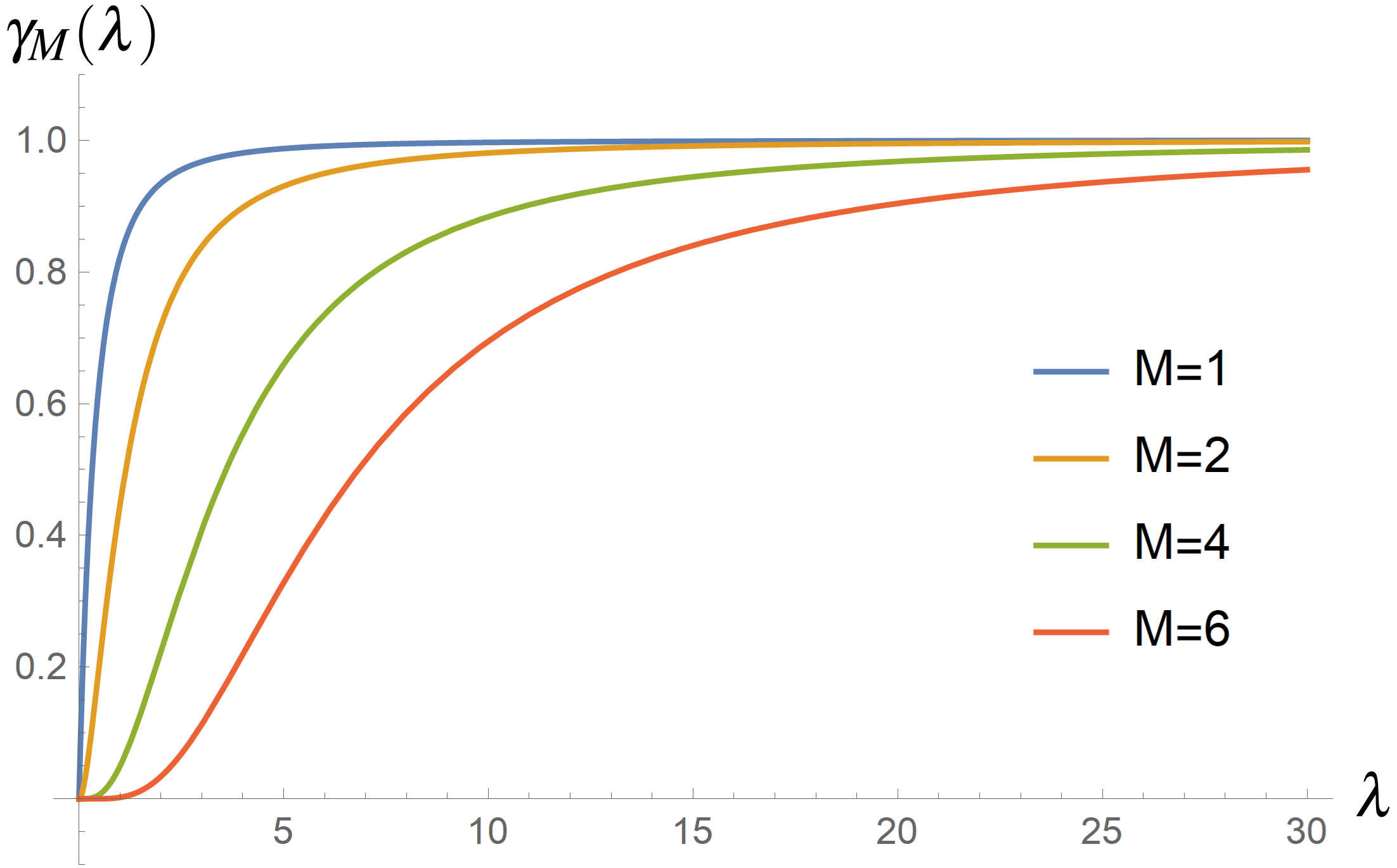}
\caption{$\gamma_{M}(\lambda)$ with different values of $M$}\label{fig3}
\end{figure}
Fig.~\ref{fig3} shows the graphs of $\gamma_{M}(\lambda)$ for $M=1,2,4,6$. It is easy to see that for any $M$ and a given radius $A$, when $\theta \to 0$, or equivalently, $\lambda\to \infty$, there is $\gamma_{M}(\lambda)\to 1$. The result (\ref{s2m}) returns to the area of the ordinary $2M$-dimensional sphere
$\frac{2\pi^{M+\frac{1}{2}}A^{2M}}{\Gamma\left(M+\frac{1}{2}\right) }$.
Furthermore, the area of a Moyal sphere is always smaller than that of an ordinary sphere with the same dimension.
When the noncommutative parameter $\theta \to \infty$, or equivalently, $\lambda\to 0$, there is $\gamma_{M}(\lambda)\to 0$.
This means that the area of the $2M$-dimensional Moyal sphere will approach to 0.

Analogous to the case of ordinary spaces, if we consider the space enclosed by the sphere to be the volume of the corresponding higher-dimensional ball, then one can see that when the noncommutative parameter $\theta \to \infty$, the volume enclosed by the Moyal sphere will also approach to 0.

Furthermore, let us consider a type of deformed Moyal space. For the $4n$-dimensional case,
the deformed Moyal product is defined as
\[ {f \star g = f \cdot e^{\frac{\mathbf{i}}{2}\sum_{i,j=1}^{4n} \Theta_{ij} \overset{\leftarrow}{\partial_i}
		\overset{\rightarrow}{\partial_j}}} \cdot g, \]
where the noncommutative parameter matrix $\Theta$ is an $4 n \times 4 n$ antisymmetric matrix
$ \Theta =\theta I_{2n}\otimes \varepsilon_2+\mu I_n\otimes
\varepsilon_2\otimes \begin{pmatrix}
	1 & 0\\
	0 & -1
\end{pmatrix}, $
and $\epsilon_{2}=
\begin{pmatrix}
	0 & 1\\
	-1 & 0
\end{pmatrix}$,  the parameter $\mu$ is a real number.
With respect to the deformed Moyal product, the variables satisfy the following commutation relations:
\begin{eqnarray}
&&\left[x_{2j-1},x_{2j}\right]_{\star}=-\left[x_{2j},x_{2j-1}\right]_{\star}=\mathbf{i}\theta,\nonumber\\
&&\left[x_{4k},x_{4k-2}\right]_{\star}=-\left[x_{4k-2},x_{4k}\right]_{\star}
=\left[x_{4k-3},x_{4k-1}\right]_{\star}=-\left[x_{4k-1},x_{4k-3}\right]_{\star}=\mathbf{i}\mu,\label{ms2}
\end{eqnarray}
where $j=1,\dots,2n$, $k=1,\dots,n$, and other commutators vanish. Obviously, when $\mu=0$, it returns to the ordinary $4n$-dimensional Moyal space (\ref{ms1}).

One can consider the variable transformations $X'=DX$, where $X=(x_{1},x_{2},...,x_{4n})$, $X'=(x'_{1},x'_{2},...,x'_{4n})$, and the transformation matrix $D$ is
\[
D=\sqrt{\frac{\theta + \sqrt{\theta^2 + \mu^2}}{2 \sqrt{\theta^2 +\mu^2}}} \left[I_{4n}+ \frac{\mu}{\theta + \sqrt{\theta^2 + \mu^2}} I_n\otimes\varepsilon_2\otimes \begin{pmatrix}
	0 & 1\\
	1 & 0
\end{pmatrix}\right].
\]
The Moyal product now becomes
\[ {f \ast' g = f\cdot e^{\frac{\mathbf{i}}{2}\sum_{i,j=1}^{4n} \Theta'_{ij} \overset{ \leftarrow}{\partial'_i}
		\overset{\rightarrow}{\partial'_j}}} \cdot g, \]
where $\Theta' = \theta' I_{2n}\otimes \varepsilon_2$, $\theta'=\sqrt{\theta^2 + \mu^2}$.
The new variables $\{x'_{1},x'_{2},...,x'_{4n}\}$ satisfy the commutative relations:
\begin{equation}
\left[x'_{2j-1},x'_{2j}\right]_{*'}=-\left[x'_{2j},x'_{2j-1}\right]_ {*'}
=\mathbf{i} \theta',
\end{equation}
where $j=1,\dots,2n$, and other commutators vanish.

Therefore, the area of the deformed $4n$-dimensional Moyal sphere (\ref{ms2}) is
\begin{align*}
	\int_{\mathbf{S}^{4n}} dvol_{\star} & = \int_{\mathbb{R}^{4n}}
	h_{\star}^{- 4n} d^{4n} x\\
	& = \int_{{\mathbb{R}'}^{4n}} h_{\ast'}^{- 4n} |D^{-1}| d^{4n} x'
	= \int_{{\mathbb{R}'}^{4n}} h_{\ast'}^{- 4n} d^{4n} x'\\
	&=\frac{(2\pi A^{4})^{2n}}{{\theta'}^{2n}}\sum_{k=0}^{\infty}\frac{C_{k+2n-1}^{2n-1}}{\left( k+\frac{A^{2}}{2\theta'}+n\right) ^{4n}}\\
	&=\frac{2\pi^{2n+\frac{1}{2}}A^{4n}}{\Gamma\left(2n+\frac{1}{2}\right) }\cdot\gamma_{2n}(\lambda'),		
\end{align*}
where $\lambda'=\frac{A^{2}}{\theta'}$.
One can see that the area of the deformed $4n$-dimensional Moyal sphere (\ref{ms2}) depends only on the radius $A$ and $\theta'=\sqrt{\theta^2 + \mu^2}$, and does not depend on $\theta$ or $\mu$ alone. Obviously, it has similar properties to the area of the normal $4n$-dimensional Moyal sphere (\ref{s2m}).

\section{Constant curvature metrics of Moyal spaces}\label{sec8}
Similar to Ref.~\cite{Eckstein}, here we will study the conformal metric with constant curvature of the 4-dimensional Moyal space, and calculate the expression of its factor $h$.
In this case, the Moyal $*$-product is defined as:
$$f(x_{1},x_{2},x_{3},x_{4})*g(x_{1},x_{2},x_{3},x_{4})=f\cdot e^{\frac{\mathbf{i}\theta}{2}(\overleftarrow{\partial_{1}}\overrightarrow{\partial_{2}}-\overleftarrow{\partial_{2}}\overrightarrow{\partial_{1}}+\overleftarrow{\partial_{3}}\overrightarrow{\partial_{4}}-\overleftarrow{\partial_{4}}\overrightarrow{\partial_{3}})}\cdot g.$$
Here we only consider the case where $f, g$ are both radial functions. Similarly, $f*g$ can be expanded to the second-order term of $\theta$,
$$
f*g =fg-\frac{\theta^{2}}{8r}\left( f''g'+f'g''+\frac{2}{r}f'g'\right) + o(\theta^{2}).
$$
If $f$ is a radial function, then $f_{*}^{-1}$ is also a radial function. Ignoring the higher-order terms of $\theta^{2}$, one can get
\begin{equation*}
	f_{*}^{-1}=f^{-1}+\frac{\theta^{2}}{4r}\left[f^{-4}(f')^{3}-f^ {-3}f'f''-\frac{1}{r}(f')^{2}f^{-3}\right]+o(\theta^{2}).
\end{equation*}
Consider the case where the scalar curvature of the 4-dimensional Moyal sphere is constant, that is
$$
S_\theta=6h_*^{2}*[\partial_{ii}(h)-h_*^{-1}\partial_{i}(h)_*^{2}-\partial_{i}(h)* h_*^{-1}*\partial_{i}(h)]*h_*^{-1}=C,
$$
where $C$ is some constant. Then there is the constant curvature equation
\begin{equation}\label{hh}
	h*[\partial_{ii}(h)-h_*^{-1}\partial_{i}(h)_*^{2}-\partial_{i}(h)*h_*^{-1}*\partial_{i}(h)]=\frac{C}{6}.
\end{equation}
Approximating to the $\theta^{2}$ term, one can assume that $h=\frac{A^{2}+r^{2}}{2A^{2}}+\theta^{2}\epsilon(r)+o(\theta^{2})=h_0+\theta^{2}\epsilon(r)+o(\theta^{2})$, where $h_0=\frac{A^{2}+r^{2}}{2A^{2}}$, then $h_{*}^{-1}=h^{-1}-\frac{4\theta^{2}A^{4}}{(A^{2}+r^{2})^{4}}+o(\theta^{2}) $. Substituting these into the constant curvature equation (\ref{hh}), after some straightforward calculations, one can obtain
$$
\frac{2}{A^{2}}+\theta^{2}\left(\frac{A^{2}+r^{2}}{2A^{2}}\epsilon''(r)+\frac{3A^{2}-5r^{2}}{2A^ {2}r}\epsilon'(r)+\frac{4}{A^{2}}\epsilon(r)+\frac{4}{A^{2}(A^{2}+r^{2})^{2}}\right)=\frac{C}{6}.
$$

If the curvature of the Moyal sphere is required to be equal to that of the classic 4-dimensional sphere $C=\frac{12}{A^{2}}$, then we have the following equation:
$$
\epsilon''(r)+\frac{3A^{2}-5r^{2}}{r(A^{2}+r^{2})}\epsilon'(r)+\frac{8}{A^{2}+r^{2}}\epsilon(r)+\frac{8}{(A^{2}+r^{2})^{3}}=0.
$$
One can see that the function $\epsilon(r)$ does not depend on the noncommutative parameter $\theta$. It is easy to find the general solution of the equation:
\begin{align*}
	\epsilon(r)=&C_{1}(r^{2}-A^{2})
	+\frac{C_{2}}{2r^{2}}\left(r^{6}-A^{2}r^{4}-17A^{4}r^{2}+A^{6}+12(r^{2}-A^{2})A^{2}r^{2}\ln r\right)\\
	&-\frac{1}{30A^{6}r^{2}(A^{2}+r^{2})}\left(A^{6}-3A^{4}r^{2}+6A^{2}r^{4}+6(r^{4}-A^{4})r^{2}\ln\frac{r^2}{A^{2}+r^{2}}\right),
\end{align*}
where $C_{1}$, $C_{2}$ are integration constants. When $C_{2}=\frac{1}{15A^{8}}$, $\epsilon(r)$ is well defined on $[ 0,\infty)$.
Furthermore, if $C_{1}=-\frac{13+12\ln A}{30A^6}$, there is $\epsilon(0)=0$.
\begin{figure}[H]
	\centering
	\includegraphics[scale=0.3]{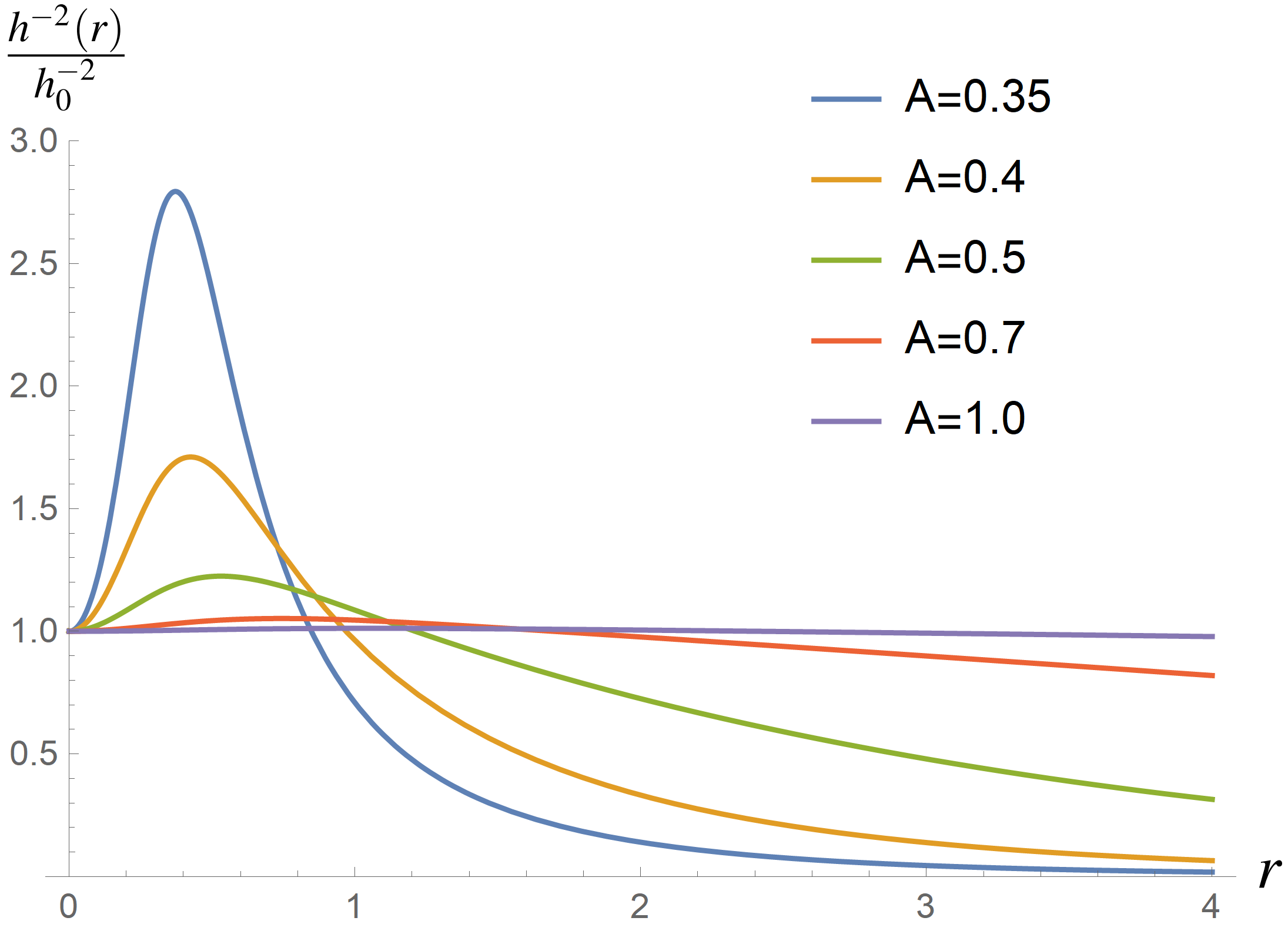}
	\caption{$\frac{h^{-2}(r)}{h_0^{-2}}$ with different radii $A$ ($\theta=0.1$)}\label{fig4}
\end{figure}
Fig.~\ref{fig4} shows the ratio of two functions $\frac{h^{-2}(r)}{h_0^{-2}}$ for different radii $A$ when $\theta=0.1$, $C=\frac{12}{A^{2}}$, $C_{1}=-\frac{13+12\ln A}{30A^6}$, $C_{2}=\frac{1}{15A^{8}}$.
One can see that for a given sufficiently small $\theta$, the smaller the radius $A$ of the sphere, the greater the deviation of the conformal metric of constant curvature of Moyal space from that of the classical case. Conversely, the larger the radius $A$ of the sphere, the closer the conformal metric of constant curvature is to the result of the classical sphere.

Similarly, one can also consider high-dimensional cases.
For any $2m$-dimensional Moyal space (\ref{ms1}), ignoring the higher-order terms of $\theta^{2}$, the approximate expansion for the Moyal product of radial functions $f, g$ is:
\begin{equation*}
	f*g=fg-\frac{\theta^{2}}{8r}\left[f''g'+f'g''+\frac{2(m-1)}{r}f'g'\right]+o(\theta^{2}).
\end{equation*}
So the corresponding inverse with respect to the star product is:
\begin{equation*}
	f_{*}^{-1}=f^{-1}+\frac{\theta^{2}}{4r}\left[f^{-4}(f')^{3}-f^{-3}f'f''-\frac{m-1}{r}(f')^{2}f^{-3}\right]+o(\theta^{2}).
\end{equation*}
Similar to the previous approach, one can assume $h=\frac{A^{2}+r^{2}}{2A^{2}}+\theta^{2}\epsilon(r)+o(\theta^{2})$. Substituting it into the corresponding constant curvature equation, it is not difficult to obtain the differential equation for $\epsilon(r)$:
\begin{equation*}
	\epsilon''(r)+\frac{A^{2}(2m-1)-(2m+1)r^{2}}{r(A^{2}+r^{2})}\epsilon'(r)+\frac{4m}{A^{2}+ r^{2}}\epsilon(r)+\frac{8(2m-1)\left[A^{2}m+(m-2)r^{2}\right]}{A^{2} (A^{2}+r^{2})^{3}}=0.
\end{equation*}

Furthermore, one can also consider the constant curvature metric of the deformed $4n$-dimensional Moyal sphere (\ref{ms2}).
The approximate expansion of the deformed Moyal product of the radial function $f,g$ is as follow:
\begin{equation*}
	f\star g=fg-\frac{\theta^{2}+\mu^{2}}{8r}\left[f''g'+f'g''+\frac{2(2n-1 )}{r}f'g'\right]+o(\theta'^{2}),
\end{equation*}
and
\begin{equation*}
	f_{\star}^{-1}=f^{-1}+\frac{\theta^{2}+\mu^{2}}{4r}\left[f^{-4}(f' )^{3}-f^{-3}f'f''-\frac{2n-1}{r}(f')^{2}f^{-3}\right]+o(\theta'^{2}).
\end{equation*}
In this case, the approximate expression of the conformal metric factor $h$ can be assumed to be $h=\frac{A^{2}+r^{2}}{2A^{2}}+\theta'^{2}\epsilon(r)+o(\theta'^{2})$. Similarly, one can obtain the differential equation of the function $\epsilon(r)$:
\begin{equation*}
	\epsilon''(r)+\frac{A^{2}(4n-1)-(4n+1)r^{2}}{r(A^{2}+r^{2})}\epsilon'(r)+\frac{8n}{A^{2}+ r^{2}}\epsilon(r)+\frac{16(4n-1)\left[A^{2}n+(n-1)r^{2}\right]}{A^{2}(A^{2}+r^{2})^{3}}=0.
\end{equation*}

From the formulas of $\epsilon(r)$ above, one can see that this type of deformed $4n$-dimensional Moyal space (\ref{ms2}) and the usual $4n$-dimensional Moyal space (\ref{ms1}) have the same solutions of $\epsilon(r)$.
So for given noncommutative parameters $\theta, \mu$, their constant curvature conformal metrics have similar properties and are also similar to those of the 4-dimensional Moyal sphere. Similarly, for given sufficiently small noncommutative parameters $\theta, \mu$, the smaller the radius $A$ of the Moyal sphere, the greater the deviation of its constant curvature conformal metric from that of the classical case; The larger the radius $A$ of the Moyal sphere, the closer its constant curvature conformal metric is to the result of the classical sphere.

\section{Conclusions}\label{sec9}
Introducing the Moyal product into the smooth function space $C^{\infty}(\mathbb{R}^{n})$ on the Euclidean space $\mathbb{R}^{n}$, one can obtain the generalized function space $(C^{\infty}(\mathbb{R}^{n}),\ast)$ with noncommutative products, and the corresponding space is the noncommutative Moyal space. If we introduce into the Moyal space a metric similar to the spherical metric of ordinary space, we will get a so-called Moyal sphere.

In this paper, we study the Moyal sphere from two different aspects.
First, with the help of the conformal metric of the sphere in ordinary space, we introduced the Moyal product in the conformal factor and obtained the corresponding spherical metric in the noncommutative sense.
We calculated the scalar curvature and area of the Moyal sphere in this case. We found that when the noncommutative parameter approaches to 0, the Moyal sphere returns to an ordinary one. As the noncommutative parameter increases, the area of the Moyal sphere will decrease. When the noncommutative parameter approaches to infinity, the area of the Moyal sphere will approach to 0. 
Analogous to the case of ordinary space, if we consider the space enclosed by the sphere to be the volume of the corresponding higher-dimensional ball, then one can see that when the noncommutative parameter approaches to infinity, the volume enclosed by the Moyal sphere will also approach to 0. In addition, we found that the total curvature integral of the two-dimensional Moyal sphere exactly coincides with the Gauss-Bonnet formula of the two-dimensional Euclidean space, and the result does not depend on the noncommutative parameter $\theta$. This implies that the noncommutative algebraic structure of the smooth function space does not change some topological properties of the two-dimensional Moyal sphere.

In the classical case, a sphere corresponds to a space with constant curvature. Therefore, we also study the conformal metric of constant curvature of the Moyal space. We obtained the approximate expression of conformal metric of the Moyal space with constant curvature in the sense of the second order approximation.
We found that for given sufficiently small noncommutative parameters, the smaller the radius $A$ of the Moyal sphere, the greater the deviation of its constant curvature conformal metric from that of the classical case; The larger the radius $A$ of the Moyal sphere, the closer its constant curvature conformal metric is to the result of the classical sphere.

In addition, we also calculated the area and constant curvature metric of a generalized deformed Moyal sphere with two noncommutative parameters, and obtained similar results. When these noncommutative parameters approach to 0, the Moyal star product will return to the ordinary product, and these curvatures and areas will return to the results of the ordinary sphere.

These results are significant for the studies of mathematical structures and physical properties of noncommutative spaces. To our knowledge, these results have
not been reported in the literatures.
Further, we will study the Gauss-Bonnet-Chern formula for the 4-dimensional Moyal sphere, and work on this direction is in progress. 

\section*{Acknowledgments}
	The authors would like to thank the anonymous referees for careful reading and valuable comments. This work is partly supported by the National Natural Science Foundation of China (Grant No. 11911530750), the Guangdong Basic and Applied Basic Research Foundation (Grant Nos. 2019A1515011703 and 2024A1515010380).

\end{document}